# Current induced anisotropic magnetoresistance in topological insulator films

Jian Wang[1,2]*, Handong Li[3,4], Cui-Zu Chang[5,6], Ke He[5], Joon Sue Lee[1], Xu-Cun Ma[5], Nitin Samarth[1], Qi-Kun Xue[5,6], Maohai Xie[3], and M. H. W. Chan[1]*

[1]The Center for Nanoscale Science and Department of Physics, The Pennsylvania State University, University Park, Pennsylvania 16802-6300, USA
[2]International Center for Quantum Materials, School of Physics, Peking University, Beijing, 100871, China
[3]Physics Department, The University of Hong Kong, Pokfulam Road, Hong Kong, China
[4]Deparment of Physics, Beijing Jiaotong University, Beijing 100044, China
[5]Institute of Physics, Chinese Academy of Sciences, Beijing 100190, China
[6]Department of Physics, Tsinghua University, Beijing 100084, China

*E-mail: juw17@psu.edu; chan@phys.psu.edu

**Topological insulators are insulating in the bulk but possess spin-momentum locked metallic surface states protected by time-reversal symmetry. The existence of these surface states has been confirmed by angle-resolved photoemission spectroscopy (ARPES) and scanning tunneling microscopy (STM). Detecting these surface states by transport measurement, which might at first appear to be the most direct avenue, was shown to be much more challenging than expected. Here, we report a detailed electronic transport study in high quality $Bi_2Se_3$ topological insulator thin films. Measurements under in-plane magnetic field, along and perpendicular to the bias current show opposite magnetoresistance. We argue that this contrasting behavior is related to the locking of the spin and current direction providing evidence for helical spin structure of the topological surface states.**

Topological insulators (TIs), materials constituting a new class of quantum matter, have recently attracted much attention in condensed matter physics[1-21]. In a TI, a finite energy gap in the bulk is crossed by the two gapless surface state branches with opposite spins which are protected from backscattering by time reversal symmetry at the Dirac points. The electron spins in the surface states are perpendicular to their momentums due to the strong spin orbit interaction. In addition to fundamental scientific interest, the topological protection of the surface states can be of interest for spintronics and quantum computation applications[2]. $Bi_2Se_3$, a narrow gap semiconductor and thermoelectric material[2,22], has been shown to be a three-dimensional (3D) TI with a single Dirac cone[1,2]. This simple Dirac cone together with the large bulk band gap (0.3 eV, equivalent to 3600 K), make $Bi_2Se_3$ to be a reference material for the 3D TIs[2-11]. In spite of existing transport measurements on 3D TIs[12-18], a key feature of the surface state, namely the locking of the spin and momentum of the conduction electrons[4, 23], has not been directly confirmed.

Recent progress in thin film growth of TIs by molecular beam epitaxy (MBE)[24-33] has made planar TI devices possible. In this paper, we measured the transport properties of the MBE-grown high quality $Bi_2Se_3$ films at low temperatures under magnetic field up to 8 T. Our single crystal $Bi_2Se_3$ films were grown under Se-rich conditions on high resistivity silicon and sapphire substrates in ultrahigh-vacuum (UHV) MBE systems[26,28]. More than 10 samples were studied and we will show in this paper data from four typical samples.

Results from the other samples show similar and consistent behaviors. Sample 1, 2 and 3 are 200 quintuple layers (QLs) $Bi_2Se_3$ films grown on high resistivity silicon substrates. Among them, samples 2 and 3 came from same film but with Hall bars patterned along perpendicular crystal axes. Sample 4 is a 45 QL $Bi_2Se_3$ film covered by 20 nm thick Se protection layer on the sapphire substrate. Under a field perpendicular to the sample, a large magneto-resistance (MR) was observed. Under an in-plane field parallel to the film sample, the MR is small and positive when the field is perpendicular to the excitation current but negative when the field is parallel to the excitation current. We attribute this contrasting in-plane MR behavior to the spin-momentum locked surface currents on TI films.

**Results**

Resistance as a function of temperature of sample 1 (200 nm thick $Bi_2Se_3$) is shown in Fig. 1a. The resistivity of the sample decreases from 125 K to 30 K. Below 30 K, the resistivity turns up with decreasing temperature, consistent with previous observations[31]. A high resolution scanning electron microscope (SEM) image of a typical MBE-grown $Bi_2Se_3$ film with a thickness of 200 QL is shown in the inset of Fig. 1a. The inset of Fig. 1b is an optical image of the Hall bar measurement structure of sample 1. The Hall resistances as a function of magnetic field are shown in Fig. 1b. The two curves at 1.8 K and 100 K are almost identical. In the low field regime, the Hall resistances are linear. Above 35 kOe, the deviation from the linear behavior is obvious. The nonlinear Hall result is likely the consequence of the presence of two channels of carriers in TIs. One from bulk and the other from the surface state[17,31,34]. The 3D electron carrier density of sample 1 is ~$2.6\times10^{18}$ $cm^{-3}$ (2D electron density: $5.2\times10^{13}$ $cm^{-2}$) and the electron mobility is 2.7 x $10^3$ $cm^2$/Vs at both 1.8 K and 100 K according to the low field Hall results.

The MR properties of sample 1 under an in-plane field are shown in Fig. 2. When the field is perpendicular to the current and crystal axis [110] (inset of Fig. 2a), the MR is positive (Fig. 2a & b) but ~10 times smaller than the MR when the field is perpendicular to the film (see supplementary information). When the in-plane field is aligned along the current direction and crystal axis [110] (the inset of Fig. 2c), other than the MR dip around zero field due to weak anti-localization effect[29,31-33], the MR is negative (Fig. 2c & d). Comparing the MR behavior in perpendicular field (see supplementary information), this negative MR effect is ~ 30 times smaller. Figures 2b and 2d show a surface plot that summarizes the MR of sample 1 as a function of both the in-plane field (from 80 kOe to -80 kOe) and temperature (from 1.8 K to 80 K). When the field is transverse to the current (Fig. 2b), the MR dip (induced by anti-weak localization[29,31-33]) at small field decreases with increasing temperature and disappears around 20 K. At larger field, the MR is positive. This positive MR effect weakens gradually with temperature. On the other hand, negative MR is clearly seen when the in-plane field is along the direction of the current (Fig. 2d). In this situation, the positive MR dip at low field due to anti-weak localization that exists at low temperature also disappears above 20 K. Near and above 50 K, a gradual positive MR is found at low field which evolves to negative MR at higher fields near and above 40 kOe.

In order to determine if this anisotropic MR phenomenon is related to the orientation of the crystal axes, a control experiment was carried out with samples 2 and 3, which are from same 200 QL $Bi_2Se_3$ film on silicon substrate. The 3D carrier density of sample 2 is $2.69\times10^{18}$ $cm^{-3}$ (2D density: $5.38\times10^{13}$ $cm^{-2}$) and the mobility is 3240$cm^2$/vs at 1.8 K. For sample 3, the 3D carrier density calculated by low field Hall resistance is $2.99\times10^{18}$ $cm^{-3}$ (2D density: $5.98\times10^{13}$ $cm^{-2}$) and the mobility is 2490$cm^2$/vs. In sample 2, the current is along the crystal axis [$\bar{1}$10] and perpendicular to [110] (see insets of Fig. 3a & b). In sample 3, the current is along [110] and perpendicular to [$\bar{1}$10] (see insets of Fig. 3c & d). Irrespective of the crystal orientation, when the field is perpendicular to the current, the MR is positive and when the field is parallel to the current, the MR is negative (see Fig. 3). Thus, we found that the transition from positive MR to negative MR is not due to the different crystal axes but the relative direction between the field and current directions.

In order to ascertain if our observations are universal, we measured a 45 QL $Bi_2Se_3$ film grown on sapphire (instead of Si) substrate covered by 20 nm thick Se protection layer (sample 4) -- see Fig. 4. The scanning tunneling microscope (STM) study of the film exhibits atomically flat terraces with 1 QL layer thick steps, which demonstrates the high crystal quality of the film (see Figure 4a). ARPES band map of sample 4 shows a single

Dirac cone with the Dirac point located at 0.135 eV below the Fermi level, which indicates a clear TI surface state of sample 4 (see Figure 4b). The Hall structure of this sample for the transport measurement is much narrower than the Hall structures of sample 1-3. The width of the Hall structure is 40 μm and the distance between two adjacent Hall bars is 400 μm (see the inset of Fig. 4c). The 3D carrier density calculated by low field Hall resistance is $3.32\times10^{18}$ cm$^{-3}$ (2D density: $1.5\times10^{13}$ cm$^{-2}$) and the mobility is 1230 cm$^2$/vs at 1.8 K. When the in-plane field is perpendicular to the measuring current, the MR is positive (Fig. 4c). This observation is consistent with that found in samples 1-3. The positive MR weakens with increasing temperature. For example, the change in resistance from zero to 80 kOe at 100 K is almost 5 times smaller than that at 1.8 K (Fig. 4c). When the in-plane field is along the measuring current, the MR is positive below 35 kOe and negative above 35 kOe (Fig. 4d) at 1.8 K. With increasing temperature, the positive MR, while weakened, is extended to higher magnetic field. At 200 K, positive MR is observed up to 80 kOe before the appearance of negative MR behavior (Fig. 4d). It is interesting that when the bias current is parallel to the field direction, negative MR is eventually observed at a sufficiently high and temperature dependent magnetic field. This behavior is qualitatively consistent with that observed in the 200 QL samples presented above.

**Discussion**

We now develop an empirical interpretation of our principal observation: namely, we find an anisotropic MR for in-plane magnetic field, independent of the crystal axes but dependent on the relative direction of the field and the current. Before seeking an explanation related to the TI surface state, we first rule out simpler explanations by contrasting our observations with the anisotropic MR occasionally found in conventional narrow band gap semiconductors that have a strong spin-orbit coupling. Such semiconductors are well-known to have spin-momentum locking in the bulk Rashba states but do not possess a symmetry-driven Dirac cone surface state. We are aware of only one such detailed study[35] that examined the MR in InSb thin films with magnetic field and current in the relative configurations identical to those presented here. Importantly, these measurements showed *no qualitative anisotropy for an in-plane field parallel and perpendicular to the current density*: while either positive or negative in-plane MR was observed with a complex dependence on film thickness and temperature, qualitatively identical MR was always seen in both in-plane field-current configurations. This suggests that the observed anisotropy in our samples is unlikely to have either a classical (skipping orbits) or quantum (weak localization) explanation based upon Lorentz force considerations alone. In the absence of any such conventional explanation, we construct a speculation by making an analogy to the anisotropic MR of ferromagnets. This anisotropy is correlated with the angle between the current direction and the magnetization – and thus -- the magnetic field[36,37]. Since the $Bi_2Se_3$ samples studied here are non-ferromagnetic single crystals, it is reasonable to speculate that our observations stem from the locking of the spin orientation relative to the momentum[19] of the conduction electrons on TI surface. In a transport measurement configuration, the collective spin polarization of the TI surface state is aligned by the current[20]. When the field is perpendicular to the current, it is parallel to the spin polarization direction. Conversely, when the field is along the current, it is perpendicular to the spin polarization of surface current. Therefore the relative direction of the current and the magnetic field may induce different MR behavior. When the field is parallel to the film, the positive MR from the classical galvanomagnetic effect is very small (Fig. 2a & b, Fig. 3a & c, Fig. 4c). The effect from the angle between the spin polarization of surface current and magnetic field direction appear to "overcome" the positive MR effect resulting in negative MR behavior in addition to the MR dip induced by possible weak-anti localization of the surface state. In some samples with suitable experimental configurations, as in our 200 QL samples, negative MR behavior is observed at very low field parallel to the bias current, while in other samples such as the 45 QL sample, positive MR is observed at low field but switches over to negative MR at a sufficiently high field. Our interpretation of the anisotropic MR in terms of spin-momentum locked surface currents is consistent with our previous findings showing that current flow strongly suppresses the superconductivity in mesoscopic electrodes in superconductor-TI film systems[23]. Both these categories of phenomena provide the first tantalizing evidence in transport studies for the presence of spin polarization at the surface of a TI.

## Methods

Single crystalline 200 nm thick $Bi_2Se_3$ films were prepared on Si (111) vicinal (3~4° offcut from the (111) to [112] direction) substrates at 200°C in a customized MBE reactor. The source fluxes of Se and Bi in the ratio of 15:1 were provided from conventional Knudsen cells. The base pressure of the system was lower than $2\times10^{-10}$ mbar. The Si(111) was thermally treated for the (7×7) reconstruction and then exposed to a Bi flux for the Bi-induced β-phase $\sqrt{3}\times\sqrt{3}$ surface structure. Prior to high temperature growth of $Bi_2Se_3$, an amorphous Bi-Se buffer layer with several nanometers thickness deposited at low temperature (~ 100 K) was employed to further accommodate the huge chemical difference between Si and $Bi_2Se_3$, facilitating the subsequent step-flow van der Waals epitaxy of the $Bi_2Se_3$ films at high temperature[26]. Biased by the vicinal surface of Si(111), twinning defects were suppressed therefore a unique pointed-step surface morphology can be characterized after the growth[26]. The growth rate was ~ 1 QL/min.

45 QL thick $Bi_2Se_3$ films were grown in an ultra-high vacuum with the base pressure below $1.5\times10^{-10}$ Torr. The substrates used for $Bi_2Se_3$ growth are commercial sapphire (0001) (Shinkosha Co., LTD, Japan). High purity Bi (99.9999%) and Se (99.999%) were evaporated from standard Knudsen cells. A Se-rich condition (Bi/Se flux ratio is about 1:15) was used in growth to obtain high quality stoichiometric films. The substrate temperature in growth was set at ~220°C. The details in growth and characterization of the films can be found in Ref. 28.

Hall bar with channel dimensions of 650 × 400 μm$^2$ was fabricated using standard photo-lithography method. Positive photoresist S1813 was spun at 4000 rpm for 45 seconds on $Bi_2Se_3$ film, followed by 110°C baking for 60 seconds. With a mask of Hall bar pattern, the photoresist-coated sample was exposed to ultraviolet light (365nm wavelength) with exposure power of 8mW/cm$^2$ for 7 seconds. The exposed part of photoresist was removed after 40 seconds of developing (MF CD-26 developer). Then, bared area of $Bi_2Se_3$ film with no photoresist on was wet-etched with 1 g of potassium dichromate in 10 ml of sulfuric acid and 20 ml of DI water. The expected etch rate for $Bi_2Se_3$ is about 60 nm per minute, and DI water rinsing is needed.

## Acknowledgements

This work was supported by the Penn State MRSEC under NSF grant DMR-0820404, the General Research Fund (No. HKU 7061/10P) and a Collaborative Research Fund (No. HKU 10/CRF/08) from the Research Grant Council of Hong Kong Special Administrative Region and by the (Chinese) National Science Foundation and Ministry of Science and Technology of China. We are grateful to Jainendra Jain, Qingfeng Sun, Zhong Fang, Chaoxing Liu, Xiaoliang Qi, Shoucheng Zhang, Mingliang Tian, Xincheng Xie and Lin He for illuminating discussions.

## Additional information

Supplementary Information accompanies this paper.

## Competing Financial Interests Statement

The authors declare that they have no competing financial interests.

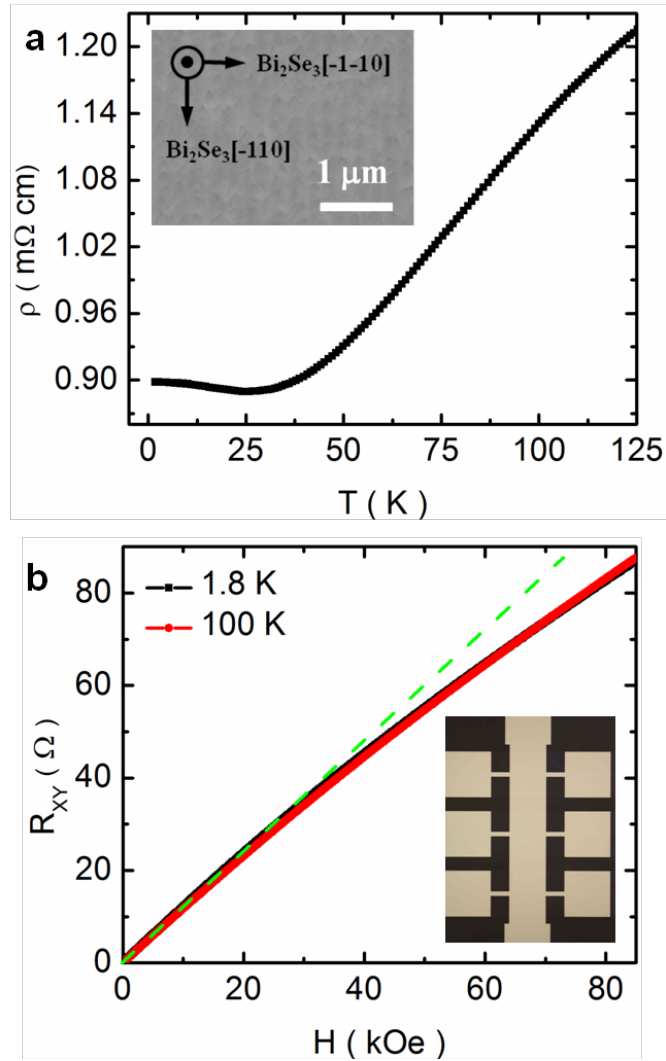

**Figure 1 | Transport property of the sample 1 (200 nm thick Bi$_2$Se$_3$ film). a**, Resistance versus temperature of the sample 1. The left inset is a scanning electron microscope (SEM) image of the Bi$_2$Se$_3$ film. **b**, Hall resistance versus magnetic field at 1.8 K and 100 K. The curves at 1.8 K and 100 K are nearly identical. The inset is the optical image of the Hall bar structure. The width of the Hall strip is 400 μm and the distance between two adjacent Hall bars is 650 μm.

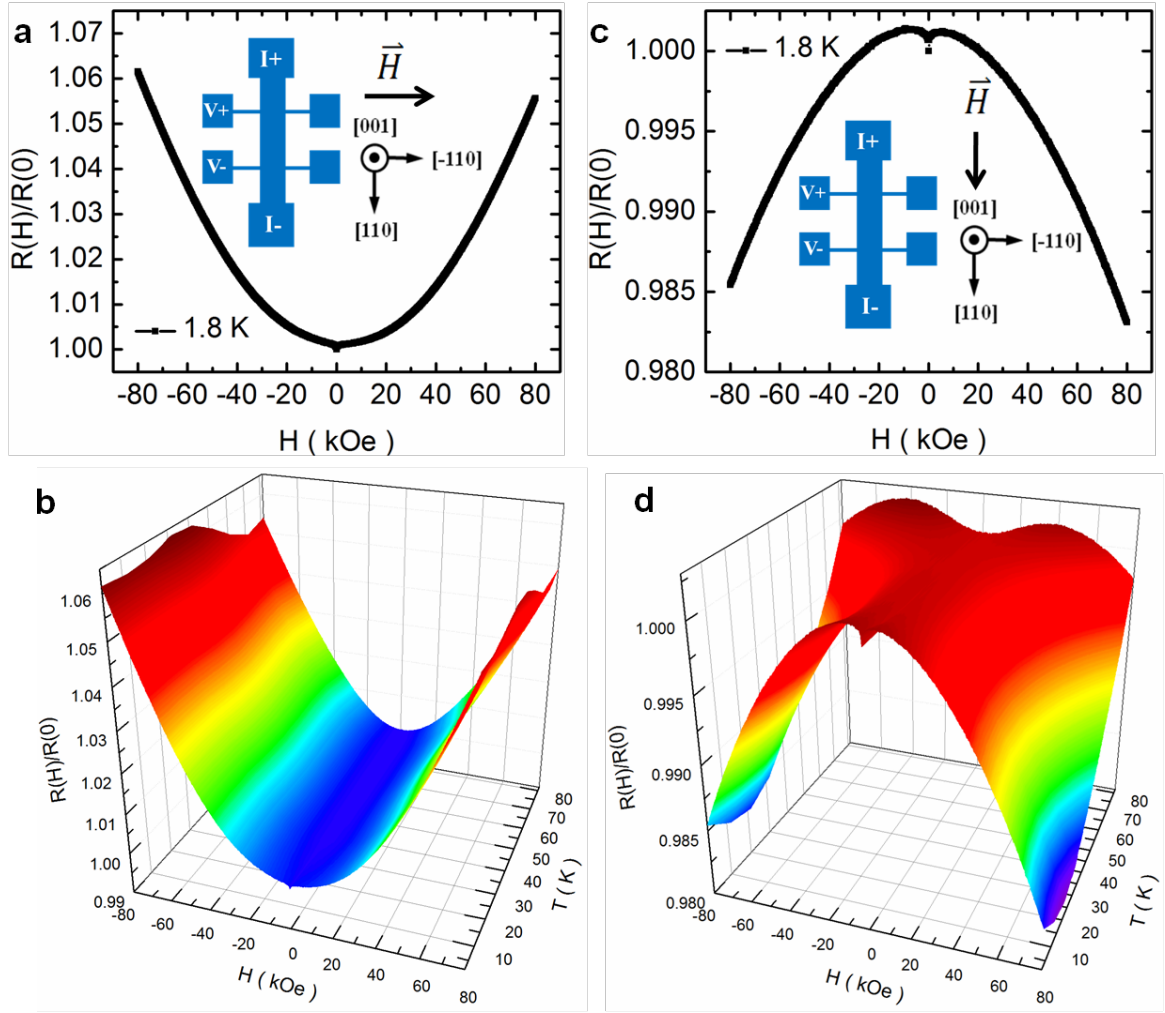

**Figure 2 | In plane magneto-resistance of sample 1 (200 nm thick $Bi_2Se_3$ film). a**, The magnetic field is perpendicular to the current direction and crystal axis [110] of sample 1 at 1.8 K. **b,** Three dimensional image of the magneto-resistance at different temperatures when the field is perpendicular to the current. **c**, The magnetic field is parallel to the current direction and crystal axis [110]. **d,** Three dimensional image of the magneto-resistance at different temperatures when the field is parallel to the current.

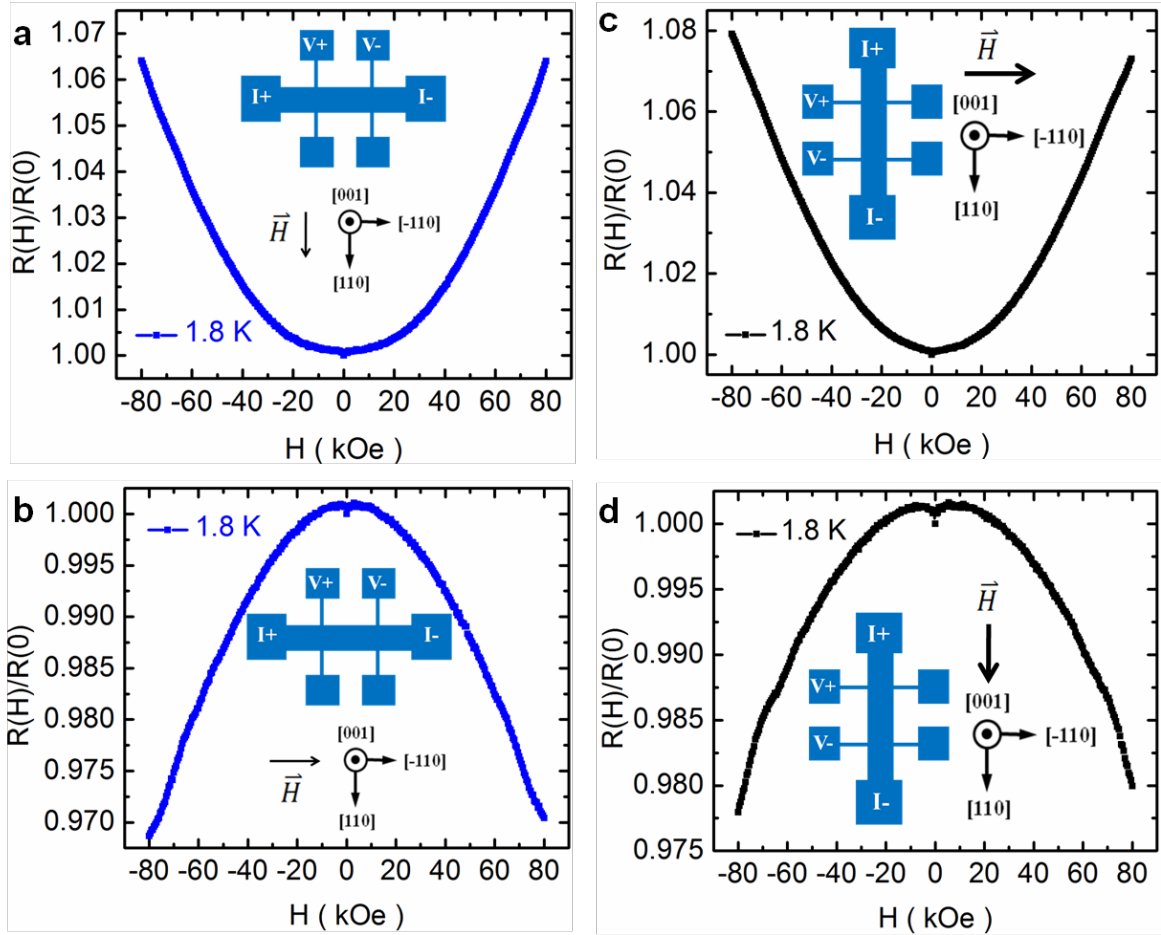

**Figure 3 | In plane magneto-resistance of sample 2 and 3 (these two samples are from the same 200 nm thick $Bi_2Se_3$ film). a**, The magnetic field is perpendicular to the current and crystal axis $[\bar{1}10]$ of sample 2 at 1.8 K. **b**, The field is parallel to the current and crystal axis $[\bar{1}10]$ of sample 2. **c**, The magnetic field is perpendicular to the current direction and crystal axis [110] of sample 3. **d**, The magnetic field is parallel to the current direction and crystal axis [110] of sample 3.

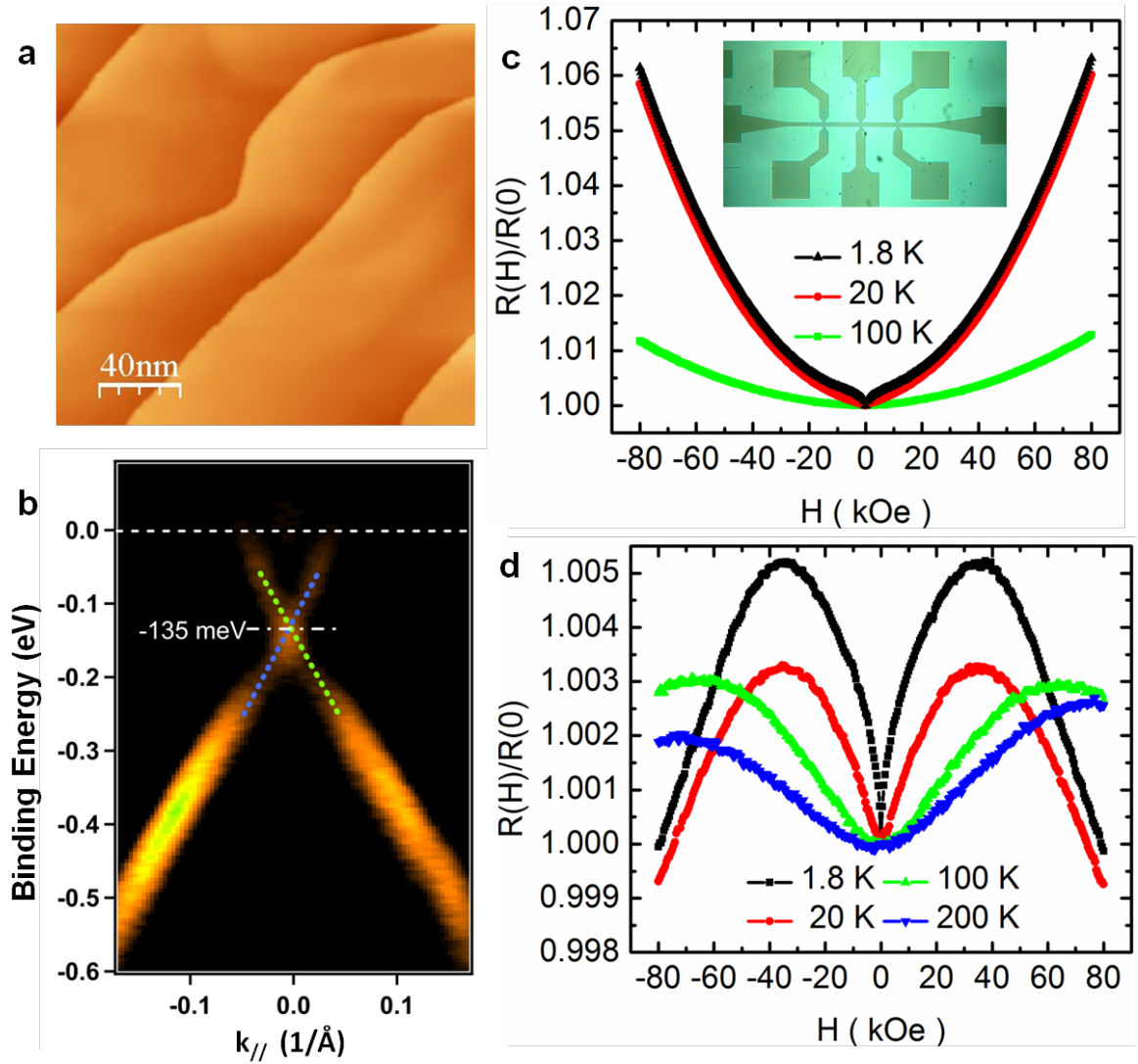

**Figure 4 | Results from sample 4 (45 nm thick $Bi_2Se_3$). a,** A typical scanning tunneling microscope (STM) image of sample 4. **b,** Angle-resolved photoemission spectroscopy (ARPES) data for the surface state of sample 4. **c,** In plane magneto-resistance at different temperatures when the field is perpendicular to the current. The inset is an optical image of the Hall bar measurement structure. The width of the Hall strip is 40 μm and the distance between two adjacent Hall bars is 400 μm. **d,** In plane magneto-resistance at different temperatures when the field is parallel to the current.